\documentstyle[prb,aps,twocolumn]{revtex}
\large
\begin{document}
\draft
\twocolumn[\hsize\textwidth\columnwidth\hsize\csname
@twocolumnfalse\endcsname

\title{\Large Magnetic Resonance of the  Intrinsic Defects of the
Spin-Peierls Magnet CuGeO$_3$ }

\author{ A.I.Smirnov, V.N.Glazkov}
\address{P.L.Kapitza Institute for physical problems RAS, 117334
Moscow, Russia}
\author{ L.I.Leonyuk, A.G.Vetkin,}
\address{Moscow State University, 119899 Moscow, Russia }
\author{R.M.Eremina}
\address{ E.K.Zavoisky Physical Technical Institute
 RAS, 420029 Kazan, Russia}

\maketitle

\begin{abstract}
\widetext
\leftskip 54.8pt
\rightskip 54.8pt

 The magnetic resonance of the pure monocrystals of  CuGeO$_3$ is
 studied in the  frequency range 9-75 GHz and in the temperature
 interval 1.2-25~K.

The splitting of the ESR line into several spectral components is
observed below 5~K, in the temperature range where the magnetic
susceptibility is suppressed by the spin-Peierls dimerization.
The analysis of the magnetic resonance signals
allows one to separate the  signals of
the $S=$1/2  and $S=1$ defects of the spin-Peierls phase.
The values of the $g$-factor, corresponding to these signals
are close to the characteristic   values  of the Cu$^{2+}$-ion
The additional line of the magnetic resonance is characterized by the
strong anisotropy of the $g-$factor and by the threshold-like
increase of the microwave susceptibility when the microwave power is
increasing. This threshold-like increase of the susceptibility is
observed both at the resonance value of the  magnetic field and at the
wings of the resonance line.  These signals are supposingly
attributed to two types of the planar magnetic defects, arising at the
boundaries of the domains of the spin-Peierls state with the different
values of the phase of the dimerization.

\end{abstract}



]

\narrowtext

\section{Introduction}

The inorganic compound CuGeO$_3$ possesses the magnetic and
crystallographic properties of the spin-Peierls crystals and is
intensively studied during the last years by different methods. The
magnetic structure of this crystal is based on the 1D-chains of
S=1/2 Cu$^{2+}$ ions, elongated along the $c-$ axis of the orthorombic
crystal.
The drop of the magnetic susceptibility is found below the
temperature of the spin-Peierls \cite{Hase}  transition. This drop of
the susceptibility is  accompanied by the lattice period doubling in
the directions $a$ and $c$ \cite{Hirota,Regnault}. The  change of the
magnetic properties is provided by  the formation of dimers of
magnetic atoms, being placed closer than in the initial state, with the
exchange integral being larger than that in the state above the
transition point.  The ground state of the spin-Peierls crystal is
singlet, and the excited states are separated by an energy gap. The
magnetic susceptibility is to be equal to zero at absolute zero. The
values of the gap and of the spin-Peierls transition temperature depend
on the variation of the exchange integral along the dimerised chain
\cite{Bray,Bulaevsky}.  Note  that the undimerised  chain of the
$S=$1/2 spins  with the Heisenberg antiferromagnetic exchange has the
gapless spectrum, and the ground state is not the Neel state
\cite{Bethe}.

 The described transition results from
the  instability produced by the interaction of the 1D spin chains with
the 3D elastic lattice of the crystal.
Spin chains construct a quasi-one-dimensional antiferromagnet, but the
reconstruction of the crystal at the transition  is of three
dimensional character and the dimers  are placed in an ordered
sublattice. The displacements of copper ions are directed along the
$c$-axes and the rotations of the oxygen octahedra surrounding the
copper ions occur in the $ab$-plane \cite{Hirota}. The
displacements of copper ions of the neighboring chains are
opposite-phase-correlated, i.e. the dimers
coincide after the translations by vectors ${\bf a+c}$ or ${\bf
b/2+c}$. Here ${\bf a,b,c}$ are the primitive translations of the
undimerized phase.  The period of the Cu-sites displacement  along the
$b-$direction is equal to a half of the period of the lattice in the
high-temperature phase, because there are two Cu-ions per primitive
cell, separated by the translation ${\bf b}/2$. The octahedra rotations
are correlated in an analogous way, the octahedra coincide after the
translations mentioned. Thus the reconstruction of the lattice is
accompanied by the dimerization of copper atoms along the $c-$ axis and
of the oxygen atoms along $a$ and $b$-axes.  The displacements
$\delta z_{klm}$
of the
$Cu^{2+}$-ions relative to the undimerised lattice
may
be described by the relation:

\begin{equation}
 \delta z_{klm} = \xi \cos[ ((k+  l +  m)) \pi +\psi]
\end{equation}

Here $\xi$ is the amplitude of the displacement, $k,l,m$ are the
coordinates of the Cu-ions respective to some reference ion in the
units of$a,b/2,c$  in the  system of coordinates attributed to the
crystal axes $a,b,c$.  The phase $\psi$ is  the phase of dimerization,
it can take one of two values:  0 or $\pi$ following \cite{Khomsky}.
The state of the crystal is doubly degenerated with respect to this
parameter.

The main characteristics of the spin-Peierls state in
CuGeO$_3$, are as follows \cite{Regnault}:
the temperature of the transition  $T_{SP}$ = 14.2~K,
the intrachain exchange integral
$J_c$ is
10.6 meV, the energy gap at absolute zero
$\Delta$ = 2~meV, the relative variation of the exchange integral in
the dimerised chain $\delta$=0.042. The interchain-to-intrachain
exchanges ratio are $J_b/J_c = $0.11, $J_a/J_c$ = -0.011.
The maximum displacement of the Cu-ions is 0.007~$A^o$.

The magnetic susceptibility data and the results of the studies of
the crystallographic and magnetic  structure and the investigations
of the excitation spectra of CuGeO$_3$  agree mainly with
the concept of the spin-Peierls transition
(see, e.g. \cite{Regnault}).

The freezing out of the magnetic susceptibility of the real samples
   does not occur for all the 100 per cents. The magnetic
   susceptibility of the best samples vanishes approximately  for
   10 times, the susceptibility takes the minimum value at 5~K and
   demonstrates an increase at the further diminishing of the
   temperature. The residual susceptibility might be ascribed to the
   dangling ends of the chains or to the impurities.  There are also
   other suggestions, given below in this paper.

The insertion of the defects in the lattice or the doping of the
magnetic subsystem results in the lowering of the spin-Peierls
   transition temperature, and in the 3D
   long range antiferromagnetic order at more lower temperatures
   \cite{Coad,Regnault2}.  After the doping by  0.07\% Si  or
   by 2\%Zn the Neel temperature is 4~K . The remarkable feature of the
   antiferromagnetic ordering stimulated by the  impurities is  the
   coexistence of the Neel and spin-Peierls states. The mean
   spin-per-site value  is of several tenths of the nominal value.  The
   suppression of the dimerization in the vicinity of the defect is the
   reason for the transformation of the nonmagnetic spin-Peierls state
   into the  antiferromagnetic state \cite{Khomsky,Fukuyama}.  The
   absence of the dimerization leads to the antiferromagnetic
   correlation of several spins  around the defect along the chain
   and in the perpendicular directions  because of the exchange
   interactions.  The average value of the spin projection
   diminishes with  moving   away from the defect.  The
   correlated areas  of the neighboring defects overlap, providing the
   long range magnetic order. At a nonzero temperature the long range
   order is destroyed when the energy of thermal fluctuations is large
   enough to damage the correlation of the neighbor defects.

    The long range magnetic order was not observed
downto 1.2~K for pure samples which showed the diminishing of the
susceptibility below $T_{SP}$ for more than 10 times.
 Nevertheless  the pure  crystals of  CuGeO$_3$ demonstrate unusual
 magnetic  properties at low temperatures.
 The  low temperature study of the electron spin resonance (ESR)
 \cite{Honda,Smirnov1}
 showed that the ESR spectrum  is complicated by splitting into
 several components of the unknown nature.
 The effect of the electric field influence on the
 magnetic susceptibility  in the temperature range of
 the residual susceptibility \cite{Smirnov2} also indicate the unusual
 state of the crystal at low temperatures.
 The residual susceptibility  was supposed to be
 provided by the magnetic defects which arise together with  the
 lattice defects  at the temperature T$_{SP}$, these defects being the
 boundaries of the spin-Peierls crystallites. The crystallites differ through
 the different values of the phase of the dimerization.  Thus the
 residual susceptibility may arise in a rather pure crystal and may
 exceed the susceptibility of the paramagnet with the  number of
 magnetic ions equal to the number of defects in the high temperature
 phase.

 The aim of the present paper is the study of the ESR spectra of the
 pure monocrystals of CuGeO$_3$ for the determination of the structure
 of the magnetic defects  of the spin-Peierls phase.

\section{Experimental techniques and samples}

The crystals of CuGeO$_3$ were grown from the high purity components
by means of the spontaneous crystallization from the flux melt at the
slow cooling. The velocity of the crystallization was  10$^{-3}$cm/hour.
The test of the impurities content was performed by means of the
activation analysis and by atomic plasma spectroscopy (ICP/APS). The
concentrations of the impurities  Fe, Ni, Mn, Co did  not exceed
10$^{-4}$ per Cu-ion.

The magnetic impurities and defects provide the residual magnetic
susceptibility in the spin-Peierls crystal. Therefore the quality
of the crystal may be characterized by the ratio $Q$ of the
susceptibility at 15~K to the minimal value of the susceptibility
measured at 5~K. The smaller is the amount of the defects in  the
spin-Peierls crystal, the higher is the value of the quality factor
$Q$.  For the main set of our samples $Q=$20.

Samples from other growth procedures were also studied
 for the comparison of the samples with the
different content of the different defects. ESR spectra of the samples
used in \cite{Smirnov1} (sample N2) were taken.  This sample was
prepared by the floating zone method and contained the impurity of Fe
in the amount of 10$^{-3}$.  The quality factor of this sample is 7.
A set of samples was grown from the same initial materials as
the main set but at the more rapid crystallization rate  for the
comparison of the samples with different concentrations of the
structure defects. The crystallization rate for these samples was
1cm/hour and the value of $Q$ was 6.  The samples grown  at the rate of
the crystallization 6  cm/h had the value of $Q$=3.
For the investigation of the  impurities influence on the ESR signal
the samples doped with Ni of the composition
Cu$_{0.995}$Ni$_{0.005}$GeO$_3$, were grown at the rate 1~cm/h.

 The lines of the magnetic resonance were taken as the dependencies
of the  microwave  power transmitted through the cavity containing
the sample vs the  magnetic field   in the frequency range
18-75~GHz. The spectrometer with the
modulation of the magnetic field was used at the frequency 9~GHz and
the field-derivatives of the magnetic resonance lines were recorded.
The measurements were done in the temperature interval 1.2-25~K in the
magnetic fields up to 60 KOe.

\section{The spectrum of the magnetic resonance}

At the temperatures above and in the vicinity of $T_{SP}$ the ESR
spectrum of CuGeO$_3$ consists of a single line. As described in
\cite{Honda,Smirnov1},  this line broadens at the diminishing of the
temperature and further the spectrum splits at 5~K,  there arise  four
strong and several weak  lines at lower temperatures.

The low temperature ESR signals are relatively weak. Using the known
value of the molar susceptibility of CuGeO$_3$ at $T$=15~K and the
value of $Q$-factor one could estimate the effective concentration of
the paramagnetic defects which could be responsible for the ESR signal
of the observed intensity. The main set samples integral
intensity at 5~K  is 1.0$^.$10$^{-3}$ of the intensity
of the electron spin resonance of the paramagnet with one electron spin
per  Cu-ion.

The characteristics of the magnetic resonance are shown in figures 1-6.

The evolution of the lineform with temperature and the transformation
of one single line into four lines  is shown on the Fig.1.
One could see here the part of the integral
intensity of the broad line is splitting and forms a new line on the
right side of the main line at T=3.5~K, the main line is also splitting
into three components.

Fig.2 shows the records of the magnetic resonance spectra at
${\bf H}  \parallel c$  at different  frequencies
 taken at the  temperature 1.3~K.
The main four lines
are marked here by the numbers
1,2,3,4.
The weaker lines are marked by the letters
$\alpha,\beta\,\gamma,\epsilon,\nu$. The dependencies of the resonant
frequencies on the magnetic field $f_i(H)$ for ${\bf H} \parallel c$
are illustrated on Fig.3. The subscript $i$ corresponds to one of the
lines marked by numbers and letters on Fig.3. Three
close lines 1,2,3 has frequency-field dependencies in the form of
three parallel straight lines. The middle straight  line of the
resonance 2 passes through the origin of coordinates. The
frequency-field dependence of the  resonance line 4   is a straight
line with another slope and also passes trough the origin of the
coordinates. The frequencies $f_{1,2,3,4}$ don't depend on the
temperature in the range 1.3-4~K. The data shown on Fig.3 and the
measurements at ${\bf H} \parallel a$, ${\bf H} \parallel b$ show that
the dependencies $f_{1,2,3,4}(H)$  for the principal orientations of
the magnetic field and within the frequency interval 9-75~GHz are
following:

\begin{equation}
f_{i}(H_\kappa)=\frac{\mu_B}{2\pi \hbar} g_{i\kappa}H_\kappa +
d_{i\kappa}
 \end{equation}

Subscript  $\kappa $ denotes one of the field directions along the axes
$a,b$ or $c$. The values of  $g-$factors  $g_{i\kappa}$ and of the
constants $d_{i \kappa}$ are given in the Table~1.
The nonzero values of $d_{1 \kappa}$, $d_{3 \kappa}$ describe the
splitting of the magnetic levels at the zero field. The zero field
splitting has the maximum value for ${\bf H} \parallel b$ and vanishes
at ${\bf H} \parallel a$.

\vspace{5mm}

{\bf Table 1}

\begin{center}
\begin{tabular}{|c|c|c|c|c|c|c|} \hline
$i$ & $g_{ia}$ & $g_{ib}$ & $g_{ic}$ & $d_{ia}$, GHz &
 $d_{ib}$,GHz & $d_{iá}$,GHz    \\
\hline
$1$ & 2.17 & 2.26 & 2.10 & 0.0 &
 -1.7 & 1.15    \\
\hline
$2$ & 2.17 & 2.26 & 2.10 & 0.0 &
 0.0 & 0.0    \\
\hline
$3$ & 2.17 & 2.26 & 2.10 & 0.0 &
 2.05 & -0.95    \\
\hline
$4$ & 1.82 & 1.86 & 1.43 & 0.0 &
 0.0 & 0.0    \\
\hline
\end{tabular} \end{center}

  The representation of the spectrum  in the field range covering the
  lines 1,2,3,4 in the form of four lorentzian lines revealed the
  presence of the fifth line which has the $g-$factor of about
  2.0 and  the linewidth of about 600~Oe.   The
  presence of this fifth line (we note it by the number "0") is obvious
  e.g. at the Fig.2 where the up and down displacements from zero
  line of the 9GHz-record of the derivative are observable at the
  resonant fields of the resonances 1,2,3.  These displacements
  correspond to the presence of a wide line in addition to lines 1,2,3,4.
  The intensity of the line "0" is about 0.07 of the integral ESR
  intensity at 1.3~K for the sample with $Q$=20.  For the sample with
  $Q$=6  the intensity of the line "0" is much larger and exceeds for
  10 times the total intensity of lines 1,2,3,4.

The dependencies of the resonant field  on it's orientation are given
in the Fig.4. The lines 1 and 3 exchange their positions so that the
frequency diffferences $f_{1,3}-f_2$ change their  signs. The rotation
of the field from the  $c-$ direction  to $a-$direction
results in the  merging  of the lines 1,2,3  into one
line.  The resonance line 4 has a strongly anisotropic $g-$factor with
the variation between the limits 1.43 and 1.86 depending on the
orientation of the  magnetic field with respect to crystal axes.

The line $\alpha$ was observable  only on the frequencies 9.1 and 9.4
GHz. The difference of the resonance field values for these frequencies
shows that this line has the zero frequency in zero field and the
$g-$factor value of 5.4 at ${\bf H} \parallel c$. The value  of the
$g-$factor  of the line $\beta$ doesn't depend on the
orientation of magnetic field and is equal to 4.21. This value is
typical for the Fe$^{+}$ - ion \cite{Altshuler}

Fig.5 shows the temperature dependence of the integral intensity of
the ESR spectrum and of the intensity of lines  3 and 4  on the
frequency 9.4 GHz.
The dependencies of the  linewidths on temperature are shown in Fig.6.
At the temperature 5~K, where the splitting into four lines occurs,
the peak of the linewidth for all lines takes place.

The comparison of the  ESR lines of the samples of different quality
is presented on Fig's. 7,8 and in the Table 2. For the samples with the
smaller $Q$-values the  intensity of the line "0"  is larger. This
fact is illustrated in Table 2 by the  ratio $\sum_{1-4}/I_0$  of
the total intensity of lines 1,2,3,4 to the intensity of the line "0"
at $T$=1.3~K. Besides the data of the present work the data from the
paper \cite{Honda} are also given. For the low-$Q$ samples the
intensity of the line "0" becomes larger, and the lines 1,2,3
broaden or disappear as observed in  crystals grown at the higher
crystallization rate.  The line 4 is also broadened and at the same
time   enlarged in the intensity.  The comparison of the samples from
the main set and of the sample N2 from \cite{Smirnov1}  shows
that the low-$Q$ sample has the stronger line 2.  The lines 1 and 3
have the same intensity and the linewidth as in the pure sample. The
line $\alpha$ is of the same intensity as in the most pure sample, and
the line $\beta $ is more intensive according to the data of the
Fe-concentration analysis.

The Fig.8 shows that the smaller is the crystallization rate the
smaller is the intensity of the line 4. This  observation is made for
the samples grown from the same initial components and confirms the
arising of line 4  from the structure defects within the magnetic
subsystem of Cu-ions rather than  from  impurity
atoms.

\small

{\bf Table 2}

\begin{center}
\begin{tabular}{|c|c|c|c|c|c|c|c|} \hline
N & $v_{cr}$, cm/h & $C_{Ni,Co,Mn}$ & $C_{Fe}$ & Q &
 $\sum_{1-4}/I_0$ & $T_{SP}$ &   \\
\hline
1 & 10$^{-3}$ & $<$10$^{-4}$ & $<$10$^{-4}$ & 20 &
 15 & 14.5 &  \\
\hline
2 & 1 &  $<$10$^{-4}$ & $<$10$^{-4}$ & 6 &
 0.1 & 14.0 &  \\
\hline
3 & 6 & $<$10$^{-4}$ & $<$10$^{-4}$ & 3 &
 0.01 & 13.0 &   \\
\hline
4 & 0.1 & $<$10$^{-4}$ & 5$^.$10$^{-4}$ & 7 &
 1 & 14.5 & [12]   \\
\hline
5 & unknown  & unknown & unknown  & 100 &
 20 & 14.5 & [11]   \\
\hline
6 & 1 & 5$^.$10$^{-3}$ (Ni) & $<$10$^{-4}$ & 2 &
 & 12.5 &    \\
\hline
\end{tabular} \end{center}

\normalsize

\section{ Nonlinear magnetic  resonance }

    The resonant microwave magnetic susceptibility appeared to be
    power-dependent for the line 4. Fig.9 shows the lines of the
    magnetic resonance absorption on the frequency 20.2 GHz at
    different levels of the incident microwave power.  When the
    microwave power exceeds some threshold value the enlarging of
    the susceptibility occurs and the intensity of the line 4 becomes
    larger than the total intensity of lines 1,2,3. The lineform
    becomes asymmetric, the additional absorption on the left wing
    is more elongated then on the right wing.

The dependence of the imaginary part of  the susceptibility  on the
     microwave power is shown on Fig.10 for the case of the resonance
     field value and for the wing of the resonance
     curve.   This dependence demonstrates the threshold for the
     enlarging of the susceptibility. The  threshold power, marked by
     the arrow on the figure, corresponds  to the incident
     microwave power of  about 1 mW and to the absorption in
     the sample of about 100$\mu$W.  The strength of the microwave
     magnetic field on the sample is about 0.1Oe.

The effect of the nonlinear enlarging of the imaginary part of the
microwave  magnetic susceptibility has the maximum value at the
frequency about 20 GHz. At the frequencies 18 and 23 Ghz this effect
also  is present but the nonlinear enlarging of the susceptibility at
the comparable power is approximately 3 times smaller. For other
frequencies (the closest of them is 26 GHz)  the nonlinear enlarging
of the susceptibility is not larger then the noise level.

\section{Discussion}

  \subsection{Temperature evolution of the lineform}

   The temperature evolution of the magnetic resonance lineform
   i.e. the transformation of the single narrow line into the single
   wide line and then into four narrow  lines  arisen from this wide line
   may be explained by taking into account the exchange interaction of the
   paramagnetic defects of the spin-Peierls phase with the thermally
   activated triplet excitations \cite{Honda}.
   The   exchange frequency is
   determined as the  product of the exchange integral expressed in the
   frequency units by the  relative concentration of the triplet
   excitations \cite{Chesnut1}.
   When the exchange frequency  is
   greater than  the difference of the frequencies of
   the different resonance lines,
   only single line with some average frequency should be observed (the
   effect of the exchange narrowing). By lowering the temperature  the
   concentration of triplets drops and when the exchange frequency passes the
   value of the order of  frequencies difference the separate
   lines arise which are narrowing at the further freezing out of the
   triplet excitations \cite{Chesnut2}.

   This scenario of the lineform evolution was observed in the TCNQ organic
   crystals, having the nonmagnetic ground state and the triplet excitations
   states \cite{Chesnut2}.
   In the pure crystals of this substance the line widening and splitting into
   two lines were  observed. These two lines corresponded to the
   triplet excitations with the effective spin $S=1$ in the crystal
   field. The line corresponding to the residual defects with the
   effective spin $S=1/2$ was found in the irradiated crystals of TCNQ
   \cite{McConnell}.  The two lines corresponding to $S=1$ vanish in
   intensity with temperature.  For  the  crystals containing defects
   the lines attributed to $S=1/2$ and $S=1$  do not disappear at the
   temperature diminishing and show the temperature dependence of the
   intensity like the intensity of a paramagnetic sample ESR.  It was
   shown in \cite{Honda}, that the linewidth of the ESR in CuGeO$_3$
   follows the temperature dependence provided by the described
   mechanism  in the range of the rapid (above 5~K)  as well as in the
   range of  the slow (below 5~K) exchange.

  \subsection{Effective spin and the origin of the defects}

  In the  temperature range below 4~K the susceptibility is
  suppressed by the transition into the spin-Peierls phase, the ground
  state of this phase is singlet and nonmagnetic, the excitation states
  being separated by the energy gap. The magnetic susceptibility and
  the intensity of the ESR signal should be exponentially close to
  zero at low temperatures. The nonzero susceptibility and the ESR
  signal are provided by the defects of the spin-Peierls phase.
  Clearly the observed ESR  signals  are attributed to different type
  of defects and further we  try to identify these defects.

The lines 1,2,3 have equal $g-$factor values, close to the $g-$factor
of the Cu$^{2+}$-ions in the paramagnetic phase of CuGeO$_3$. This fact
shows that the lines 1,2,3 are attributed to Cu-ions. The value of the
$g-$factor of the line 4 differs strongly from the Cu$^{2+}$ $g-$factor
value. One could assume this line to be provided by  impurity. But
the chemical composition   analysis  gave the limit of the
paramagnetic   impurities concentration corresponding to only a
half of the observed magnetic resonance intensity.
We recorded the ESR spectrum of the single crystal with the composition
Cu$_{0.995}$Ni$_{0.005}$GeO$_3$. The impurity ESR line of this crystal
corresponds to the g-factors
$g_a$=1.92, $g_b$=2.00, $g_c$=1.7, i.e.  it's frequency differs
sufficiently from the frequency of line 4. Thus the presence of the
nickel impurity could not explain  this signal. The magnetic
resonance signals provided by the nonmagnetic doping  do not show
$g-factor$ changes above 3\% \cite{Fronzes}.
Besides that the difference in the intensities of the line 4 of the
samples prepared at different crystallization rates from the same
components testifies the origin of this line from Cu-ions.  Thus the
results obtained demonstrate that the resonance  lines 1,2,3,4 arise
from the Cu-ions.

The relative intensity of the line 2 with respect to the
lines 1,3  is different for the samples of different quality
(see Fig.7). The ratio of the intensities of lines
1 and 3 is the same for different samples. From this fact we conclude
that the lines 1,3 are provided by the defects of certain type, and the
line 2 arises from the defects of another type. The absence of the
crystal field splitting for the line 2 shows the effective spin
$S=$1/2 of this type of defects  \cite{Abragam1}.

The zero field splitting and the angular dependence of lines 1,3 with
the exchange of their mutual positions  mean that
these lines belong to the defects with the effective spin $S=$1.
The splitting of the magnetic resonance line by the crystal field into
two lines is characteristic for spin $S=$1. The angular dependencies of
these two lines resonance fields are analogous to the dependencies shown
in Fig.5, the separation between the lines being independent on frequency
when $g\mu_BH \gg D $, here  $D$ is  the single ion anisotropy constant of
the spin-Hamiltonian. \cite{Abragam1}.
Therefore we assume that the lines 1 and 3 are concerned with the
exchange-coupled  $S=1$ pairs of Cu-ions. The ESR-line of these pairs
may be split by the dipole-dipole interaction or by the anisotropic
exchange \cite{Abragam2}. Such a splitting was observed in the
ESR spectrum of copper-acetat-monohydrat \cite{Abragam2}. The
consideration of dipole-dipole interaction or of the anisotropic
exchange is necessary because  the splitting of the spectrum of
the  pairs of $S=1/2$-ions by the single ion anisotropy is absent.

  \subsection{Intrinsic defects of the spin-Peierls phase}

We found in our experiments that the line 2 of the $S=1/2$-  and
the lines 1,3 of the $S=1$-defects have the comparable
intensities.  For the random distribution of the small number of the
defects the exchange coupled pairs  resonance  should be much less
intensive than the resonance of the isolated defects.
For an explanation of
this contradiction one should consider the structure of the magnetic
defects of the spin-Peierls crystal arising at the transition point.

  As it was described in the Introduction,  the low temperature phase
is characterized by one of the two values of the dimerization phase.
The creation of the domains (crystallites) with the different values of
this parameter is possible during the transition. On a boundary between
two crystallites the  value of $\psi$  changes from $\psi=0 $ to
$\psi=\pi$  and at least one atomic layer remains
undimerized.
The boundaries of the crystallites are usually pinned at the defects
and thus one point defect in the high temperature phase may produce the
entire plane of the magnetic defects below the
transition. The known boundaries of the antiferromagnetic
domains are examples of the plane-type defects in a
relatively perfect crystal\cite{Pisarev}.
Fig.11 shows schematically the Cu-ions in the dimerized crystal
containing two domains.  The orientation of the   plane parts of
the boundary is chosen to be  directed along the principal directions.
The boundary of the first type (I) lies in the plane $ac$ and contains
the undimerized chains oriented along $c$-axes  coupled by the weak
ferromagnetic exchange $J_a$.
The passing of this boundary violates the order of the oxygen atoms
 displacement.
  In the second type of boundaries (II), lying in the plane $bc$ (not
  shown in the Fig 11.) the undimerized spin chains are coupled by the
 weak antiferromagnetic exchange $J_b$.
 In both  cases there is a strong intrachain exchange interaction
 characterized by the exchange integral
 $J_c$.
 The violation of the order of joining of Cu-ions into pairs takes
 place in the boundaries  of the third type (III) lying in the planes
 $ab$  or in the family of the planes \{101\}. This type of boundaries
 contains the undimerized spins from different chains.
 The week exchange interactions $J_a$ and $J_b$
 are coupling the spins  in the  plane of this boundary.

The type I boundary contains the  magnetically unordered spin
chains with the strong intrachain exchange $J_c$. The week
ferromagnetic exchange acts in the perpendicular direction. The
magnetic susceptibility of these chains is suppressed by the strong
exchange $J_c$ and is of about 1/200 of the susceptibility of the same
amount of paramagnetic spins at $T=$1.5~K.
The structure defects in this boundary (vacancies cutting
the chains or the steps)  should result with a high probability in the
creation of $S=1$ exchange coupled pairs of the Cu-ions due to the
ferromagnetic exchange $J_a$ between the spins of the
neighboring chains.
The breaks of the chains in the boundaries of the type I
are probably the sources of the  exchange coupled pairs giving
the ESR lines 1 and 3.

The angular and frequency dependencies of the magnetic resonance field
    of lines 1 and 3 may be described on the base of the
 spin-Hamiltonian of the effective spin $S=1$
 \cite{Abragam1}:

\begin{eqnarray}
   {\cal H}& = g_{c}\mu_B H_{c} +g_{b}\mu_B H_{b} +g_{a} \mu_BH_{a} +
   D_{c} \hat{S}_{c}^{2} + \nonumber \\
    & D_{b}\hat{S} _{b}^{2} + D_{cb}(
   \hat{S}_{c}\hat{S}_{b}+\hat{S}_{b}\hat{S}_{c}) \label{hamilt}
   \end{eqnarray}

with the parameters  $ g_{c}$ = 2.10, $g_{b}$ = 2.26, $g_{a}$ = 2.17,
$D_{c}$ = 0.04 K, $D_{b}$= $-$0.05 K, $D_{cb}$ = $-$0.03 K

 The analogous defects in the main spin-Peierls matrix or in the
 boundaries of the types II and III remain mostly unpaired or create
 pairs with the spin $S=0$ due to the stronger antiferromagnetic
 exchanges along the axes $c$ and $b$.

 The resonance line 2 is naturally combined with the isolated breaks of
 the spin chains away from the boundaries or in the boundaries of the
 type II. These breaks produce the free $S=1/2$ spins.

 The splitting of the ESR line into three spectral
 components at lowering the temperature  was also
 observed in the organic spin-Peierls crystals\cite{Jacobs}. In this
 case the analysis of the angular- and frequency-dependencies of the
 resonant fields of theses components was not performed. The authors
 interpreted the observed components as ESR signals of different
 magnetic ions with different $g-$factors.

  The specific transformation of the ESR line in the spin-Peierls
  magnet with the splitting of the single line into several lines and
  among them a triplet was confirmed by the observation of the
  described scenario of the evolution of the lineform for the lately
  discovered second inorganic spin-Peierls crystal
  NaV$_2$O$_5$ \cite{Ueda}.

  \subsection{Magnetic clusters in the spin-Peierls matrix}

  We consider now the domain boundaries of the
  type III, having the undimerized spins from
  different chains in the plane. According to the concept developed in
  \cite{Khomsky,Fukuyama} each undimerized spin is a center of a
 region  with the
 antiferromagnetically correlated spins.
 The size of this region (soliton) is estimated theoretically as about
 7 lattice periods along the $c-$axes. The total spin of the soliton is
 1/2 and the average of the spin projection on the site diminishes with
 going away from the undimerised spin due to the dimerization (see Fig.
 12).

 These objects are magnetic clusters  within the nonmagnetic
 spin-Peierls matrix. Magnetic resonance of this clusters having the
 internal structure is an yet unsolved problem. The related problem of
 magnetic resonance  of the three-spin cluster of equal $S=1/2$ ions is
 reported \cite{Tsukerblat}.  For the  isosceles triangular cluster
 with the small deviation from the equilateral triangle  the energy
 levels are given by the relation:

\begin{eqnarray}
 E_{1,2,3,4}&= \pm \frac{1}{2} (G^2+\delta^2+g^2\mu_B^2 H^2
 \nonumber \\
 &\pm 2g\mu_BH (\delta^2+G^2\cos{^2\theta})^{1/2})^{1/2}
\end{eqnarray}

  here $\theta$  - is the angle between the  $z$ -axes of the cluster
  symmetry and the magnetic field,
  $J_{0,1} $ -exchange integrals, $\delta=\mid J_1-J_0 \mid $, $G$-
  coefficient of the antisymmetric exchange
  interaction of Dzyaloshinski-Moria. The coefficient $G$
  equals zero if the symmetry center exist between the ions
  constructing the pair. If the   energies of transitions
  between the levels (4) are small compared to $\delta, G$ then they are
  determined by the relation

\begin{equation}
\hbar\omega = \frac{g}{\sqrt{1+G^2\cos{^2\theta}/\delta^2}}\mu_BH
\end{equation}

  Thus in the low-frequency range the spectrum of the magnetic
  resonance of the triangular cluster is  analogous to the ESR
spectrum of an isolated  single ion with the strongly anisotropic
$g-$factor and with $g_z < $2.  The experiments with the organic
complex crystals containing the triades of Cu-ions revealed the
correspondence of the  static magnetic susceptibility
\cite{Bazhan}  and of the resonant properties \cite{Yablokov} to the
consideration described above, considering the nonzero value of $G$.

For the description of the linear three-spin cluster created
within the nonmagnetic spin-Peierls matrix around an undimerized spin
we take the Hamiltonian in the form:

\begin{eqnarray}
  {\cal H}& =
  J_{12}{\bf \hat{S}}_{1}{\bf \hat{S}}_{2}+
  J_{23}{\bf \hat{S}}_{2}{\bf \hat{S}}_{3}+
  J_{13}{\bf \hat{S}}_{1}{\bf \hat{S}}_{3}+
  {\bf G}_{12}[{\bf \hat{S}}_{1} {\bf \hat{S}}_{2}]
  \nonumber \\
  & \mbox{}+{\bf G}_{23}[{\bf \hat{S}}_{2}{\bf \hat{S}}_{3}]
\label{interaction}
\end{eqnarray}
here ${\bf G}_{12}$ and ${\bf G}_{23}$ are vector parameters of the
Dzyaloshinski-Moria-interaction. The energies of the two lowest states
of the cluster at the arbitrary orientation of the magnetic field are
given by the relation:

\begin{eqnarray}
  E_{1,2}&= \frac{\varepsilon_{1}+\varepsilon_{2}}{2}-
  \frac{1}{2} [(\varepsilon_{1}-
\varepsilon_{2})^{2}+{\bf G}^2+({\bf h})^{2} \nonumber \\
& \mbox{} \pm 2
((\varepsilon_{1}- \varepsilon_{2})^{2}{\bf h}^2
+({\bf h G})^{2})^{1/2}]^{1/2} \nonumber \\
&
\label{energy}
\end{eqnarray}
here $\varepsilon_{1}$ , $\varepsilon_{2}$ are
the energies of the two possible $S=1/2$ states of the cluster
in the absence of the antisymmetric
exchange and of the magnetic field:

\begin{eqnarray} \varepsilon_{1,2}& =
-\frac{1}{4}( J_{12}+J_{13} + J_{23}) \nonumber \\
& \mbox{} \pm
\frac{1}{2}\sqrt{\left[J_{23}-\frac{1}{2}(J_{12}+J_{13})
\right]^{2} + \frac{3}{4}(J_{12}-J_{13})^{2}} \label{epsilon}
\end{eqnarray}

${\bf G}$ and ${\bf h}$ are
 ${\bf G}=\frac{{\bf G}_{12}+{\bf G}_{23}}{\sqrt{3}}$,
 ${\bf h}= g \mu _B {\bf H} $,
$g$  takes the values of the corresponding components of the $g-$tensor
of the single Cu-ion.

   For the case  ${\bf h} \perp {\bf G}$
   the effective $g-$factor is approximately given by
\begin{equation}
   g_{eff} = g \left[ 1 - \frac{G_{x}^{2}+G_{y}^{2}}{2(\varepsilon_{2}
- \varepsilon_{1})^2} \right]
\label{triada}
\end{equation}

Taking $J_{23}=J_{12}$ = 10 meV \cite{Regnault}, $J_{13}$ =3.6
meV \cite{Riera}
we obtain for the observed
$g_{á}$ = 1.43  that the vector ${\bf G }$
should be  perpendicular to the plane of the fragment
CuO$_{2}$-CuO$_{2}$-Cu with the value of  $ G_{12}+G_{23} $
of about  8 meV.  Note that
${\bf G}$=0
 for the cluster with the center of
symmetry  at the middle spin.
However, as it is seen from Fig.11 there is no symmetry centers on the
sites of the undimerized spins within the boundary of the type III. The
symmetry centers disappeared due to the  distortions of the
regular dimerized pattern for the neighboring pairs of the
Cu-ions.

The estimation of the antisymmetric exchange given above is of the
order of the main exchange interaction which seems to be
nonrealistic.  Further we consider the five spin model
(Cu5-Cu4-Cu1-Cu2-Cu3). Here the number of the states with the total
spin
$S=\frac{1}{2}$ which are admixed to the main doublet  by the antisymmetric
exchange is enlarged. As a consequence the relation
(\ref{triada}) transforms in the following way:

\begin{eqnarray}
g_{eff}& = g [1 - \frac{({\bf
G_{12}}+{\bf G_{23}}+{\bf G_{41}}+{\bf G_{54}})^{2}} {3 E_{31}^{2}}
\nonumber \\
&
- \frac{({\bf G_{12}}+{\bf G_{23}} - {\bf G_{41}} - {\bf
G_{54}})^{2}}{12 E_{21}^{2}} ]
\label{penta}
\end{eqnarray}

here
 $E_{31}^{2}$ and $…_{21}^{2}$ are determined by the equations:

\begin{equation} …_{21}^{2} = [
J_{23}-\frac{1}{2}(J_{12}+J_{13})+\frac{1}{4} J_{24}]^{2}
+\frac{3}{2} (J_{12}- J_{13})^{2}
\label{e21}
\end{equation}

\begin{eqnarray}
E_{31}^{2} = &
 \{J_{23}-\frac{1}{2}(J_{12}+J_{13})- \frac{3}{4} J_{24}
 + \nonumber \\
& [(J_{23}-\frac{1}{2}(J_{12}+J_{13})  +
 \frac{1}{4}J_{24})^{2} +  \nonumber \\
& \frac{3}{2}(J_{12}-J_{13})^{2}
  ]^{1/2}\}^{2}
  + \frac{3}{2}(J_{12} - J_{13} )^{2}
  \label{e31}
  \end{eqnarray}

Note that only the antisymmetric term $G_{12}+G_{23}$
is included in (\ref{triada}) while the symmetrical contribution
$G_{12}+ G_{23} - G_{41} - G_{54}$ arises in  (\ref{penta}).
This principal difference is due to the fact that the
three-spin cluster  has the single excited state with the total spin
$S=\frac{1}{2}$ and this state is antisymmetric. There are the
symmetric as well as antisymmetric with respect to the center of the
cluster  states  among the  $S=\frac{1}{2}$ excited states of the
five-spin cluster.  Therefore both the antisymmetric and the symmetric
combinations $G_{12}+G_{23}+G_{41}+G_{54}$, $G_{12}+G_{23}- G_{41}
-G_{54}$ contribute to the change of $g-$factor.

The five spin cluster has four $S=\frac{1}{2}$ excited states.
The relation  (\ref{penta}) takes into consideration only two lower
states otherwise the formula becomes too cumbrous.
The calculation under the assumption
$G_{12}=G_{23}=G_{54}=G_{41}$
results in $g_c$=1.5 at  the modulus of $G_{12}$ equal to 2.9 meV.
  The following numerical data are used here:
$J_{12}=J_{14}=$10 meV, $J_{23}=J_{45}=$10.4 meV
(dimerized spin pairs ), $J_{13}=J_{15}=$3.6 meV.

The numerical calculations performed by the exact diagonalization of
the energy matrixes in the presence of the magnetic field including all
the excited states of the five spin cluster give $G_{12}$ = 3.0 meV.

 The average values of the spins projections of the spins of the
  cluster obtained during the process of this calculation are given in
Fig.12.
The presence of the antisymmetric exchange in CuGeO$_3$ with the
Dzyaloshinsky vector perpendicular to the $c-$direction was at first
assumed in \cite{Yamada} at the analysis of the reasons of the
high-temperature ESR line broadening.
It was noted that this assumption contradicts to the crystal
structure  of the CuO$_2$-chains reported in \cite{Hirota}.
We note here, that  the symmetry is lowered in the vicinities of the
undimerized spins contained within the domain boundaries of the type
III and the presence of the Dzyaloshinski-Moriya interaction becomes
permitted at least below $T_{SP}$.

 The  ESR line 4 is obviously to be associated with the defects of the
 last type of the Cu-ions magnetic system. The consideration of
 five-spin-cluster  given above enables one to explain the strong
 deviation of the $g-$factor from the value 2.0 and the strong
 anisotropy of $g-$ factor by taking into account the antisymmetric
 exchange interaction with the parameter  of about 0.3 of the exchange
 integral.

 Surely the above consideration of the magnetic cluster with the
 internal structure defined by the Dzyaloshinsky-Moriya interaction
give only qualitative explanation of the strong deviation of
$g-$factor.
 There are  two following contradictions of the described model to the
 experimental facts. At first, the value of $g-$factor
does not come close to 2.0 for any direction of the magnetic field
(there should be such a direction according to the model), and secondary
- the strong diminishing of $g-$factor is to be for any direction
perpendicular to ${\bf G}$ but we observe the strongest diminishing
only along $c-$axes.
 Probably these discrepancies may be ascribed to the
antisymmetric exchange of the next-nearest-neighbor-ions or to the nonparallel
Dzyaloshinsky vectors of the different pairs of ions.

The 9~GHz-ESR of the crystal of  higher  quality was reported in
\cite{Honda}. The data given there testify the $Q$-value of about
100. The ESR lines $\alpha, 1,2,3,4$ were also observed there and the
resonance fields of these lines correspond well to the fields
observed in the present paper. This fact confirms that the spectrum
consistent from a triplet line and of the cluster line is
characteristic for pure spin-Peierls crystals.

  \subsection{2D-magnet on the boundary between the domains of
  the spin-Peierls phase}

 The localized  soliton may be considered as a magnetic
 quasiatom because  it's internal structure is fixed by the strong
 exchange $J_c$.
Thus the type III boundary is a two-dimensional-magnet of these
   quasiatoms, coupled by the ferromagnetic exchange $NJ_a$ along
 $a-$direction and by the antiferromagnetic exchange $NJ_b$ along the
 $b-$axes.
Here $N$ is the effective number of spins within the quasiatom.
 Using for an estimation $N \approx 5$ we derive the exchange
 integrals between quasiatoms of about 50~K along $b$ and about
 $-5$~K along $a-$axes.
 Due to the anisotropy energy of per quasiatom-spin of $E_a$=0.5~K, the
 ordering temperature of this 2D-magnet is of about

  $T_c \approx N(J_bJ_a)^{1/2}/ln(NJ_b/E_a) \approx 3K$

 Therefore at the temperature of our experiments when
 $T\ll NJ_b$
 this planar magnet may be unordered but  strongly correlated
 magnet.
 The long-wave excitations of this magnet are
 analogous to spin waves in the antiferromagnet without anisotropy.
 \cite{MuellerBonner}. One of the branches of these excitations spectrum
 $\omega_{1 {\bf k}}$ is gap-less even in presence of the magnetic
  field.
  The second branch has the gap
  $ \omega_{20} = g_\alpha \mu_B H_\alpha$.
 The uniform high frequency magnetic field enables us to excite the
 uniform precession mode with the frequency $\omega_{20}$.
 Because of the elliptical trajectories of the spin precession caused
 by the anisotropy, the parametric excitation of the pairs of spin
 waves of the gap-less mode is possible by the decay of the uniform
 mode, when the pair meets the condition of the parametric resonance:

 \begin{equation}
 \omega_{mw}=
 \omega_{1 {\bf k}}
 +\omega_{1 {-\bf k}}
 \end{equation}

 Here $\omega_{mw}$ is the frequency of the microwave pumping.
The absorption of energy at the parametric excitation has a threshold
     in pumping power, the  flow of energy into the
 spin-wave modes  having the  resonance maximum near the frequency of
 the uniform mode (see, e.g.
 \cite{Ozhogin,Prozorova}).  The resonance is due to
 the transfer of energy via the magnetization oscillation of the
 gap-like-mode.

The parametric excitation of the spin waves of the gap-like mode
should provide the absorption of the microwave magnetic  field energy
in the field range below the resonance field of the  uniform
precession at a half frequency. We observed such absorption bands, as
  is it seen on Fig.2  and Fig.3 (line $\nu$).

The observed effect of the threshold enlarging of the susceptibility
at the resonance of the line 4 leads one to the conclusion that there
are the two dimensional planar magnetic defects inside the spin-Peierls
matrix.  The point-like magnetic defects could not provide the
nonlinear effect with the enlarging of the susceptibility, only the
saturation effect is known for the magnetic resonance of the isolated
ions, with the diminishing of the imaginary part of the
susceptibility at enlarging microwave power.
The presence of the three-dimensional magnetically correlated areas is
less probable because 3D-ordering temperature should be
greater, of the order of
   $(J_cJ_b)^{1/2}\geq$ 10K and
the 3D-dimensional order would result  in the
observable zero-field gaps for both branches of the spectrum.
This assumption is in a  contradiction with the linear and gapless
dependence $f_4(H)$.

\section {Conclusion}

The defects of the spin-Peierls phase with the effective spins 1/2 and
   1 are
  identified on the base of the analysis of
  the low temperature ESR spectra of pure crystals of CuGeO$_3$.
    Spin-1/2-defects are created by the breaks of the spin chains and
  spin-1-defects - by the exchange coupled pairs of this breakups placed
  in the boundaries of the domains of the spin-Peierls phase.

  The additional ESR signal is found revealing the two-dimensional
  magnetic defects with the long-range magnetic correlations. This
  planar objects are proposed to be the boundaries of the spin-Peierls
  phase with the different values of the dimerization phase.

\section {Acknowledgments}

\vspace{3mm}

The authors are indebted to L.A.Prozorova, V.A.Atsarkin, V.V.Demidov,
I.A.Zaliznyak, V.A.Marchenko, S.S.Sosin, L.E.Svistov and H.Benner for
valuables discussions, H.-A.Krugg~von~Nidda for taking part in the
measurements at the frequency 9 GHz, Yu.M.Tsipenyk, V.I.Firsov and
M.Fedoroff for the element analysis of the samples.

The work is supported by the Russian Foundation for Fundamental
Research (project 98-02-16572) and by the Civilian Research and
Development Foundation (project RP 1-207). The measurements on the
frequency 9GHz are made in Technische Hochschule Darmstadt, Germany
supported by Sonderforschungbereich 185.

{\bf Figure captions}

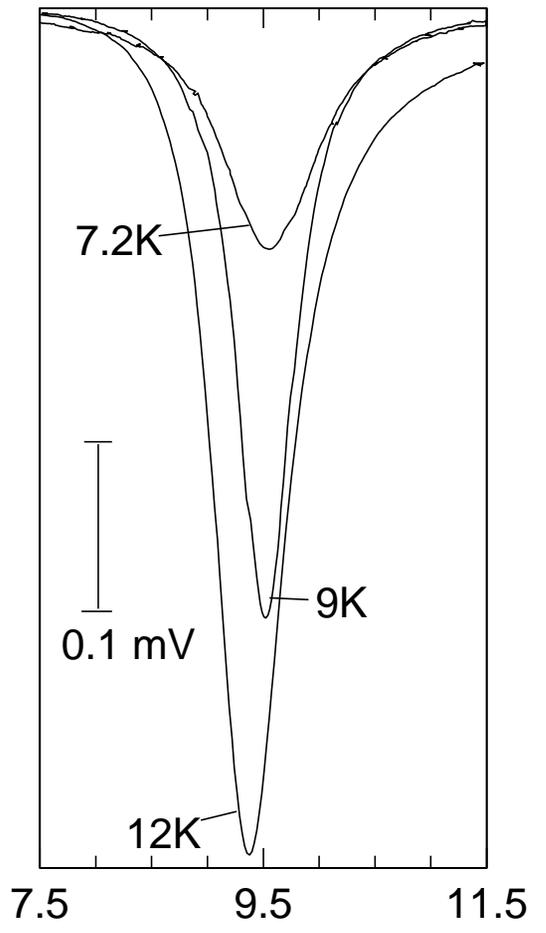
Fig.1. 26.7~GHz magnetic resonance lines taken at different
temperatures (field dependencies of the signal of the microwave power
detector)

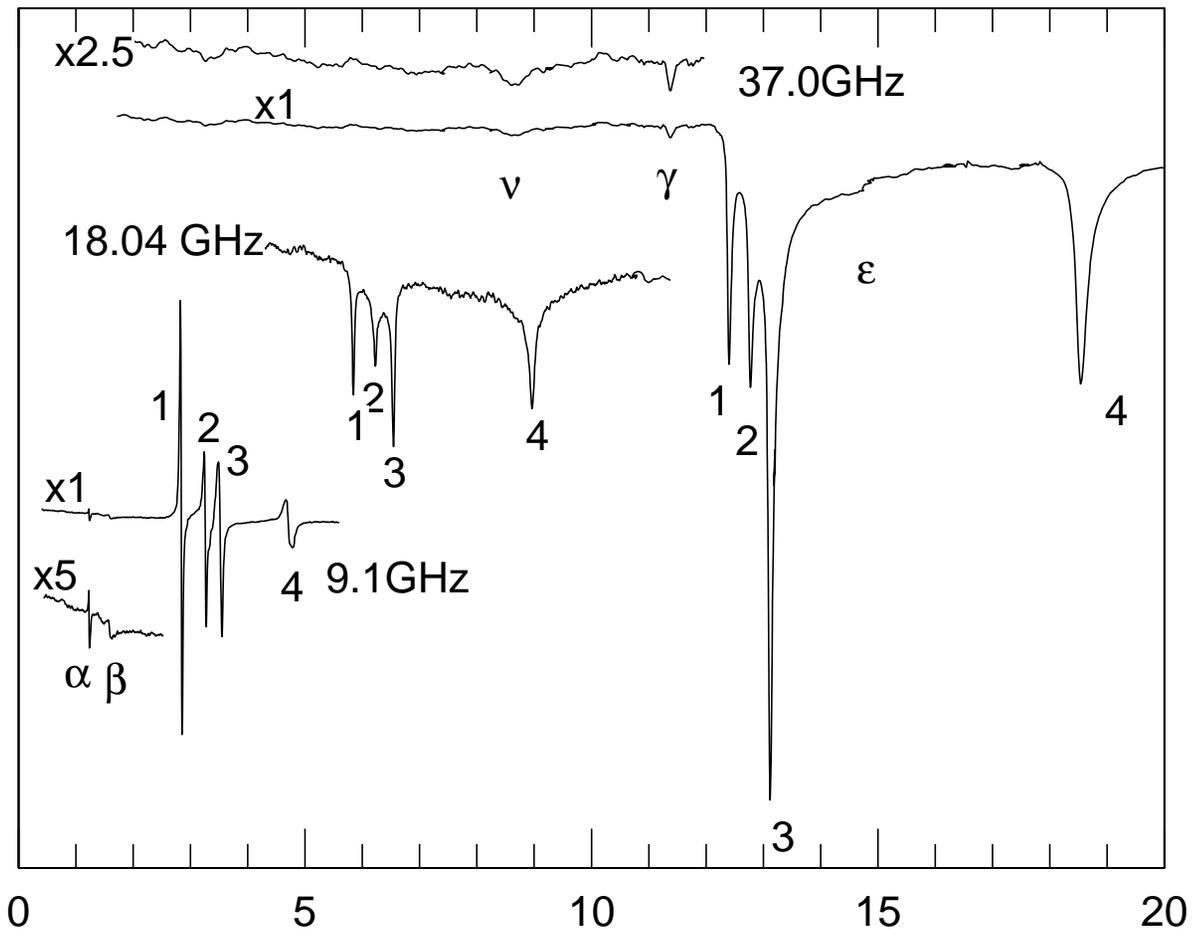
Fig.2.
Magnetic resonance lines of  CuGeO$_3$ at
${\bf H \parallel c}$, taken at the frequencies
37.0, 18.0,  9.1 GHz and  at the temperature¥ 1.2~Š. For the frequency
9.1 GHz the field-derivative of the ESR line is given.

Fig.3 ESR spectrum at
${\bf H} \parallel c$, T=1.3~K.
The deviations of the resonant frequency from the line 2 resonance
frequency
$f=2.88H$ are given.

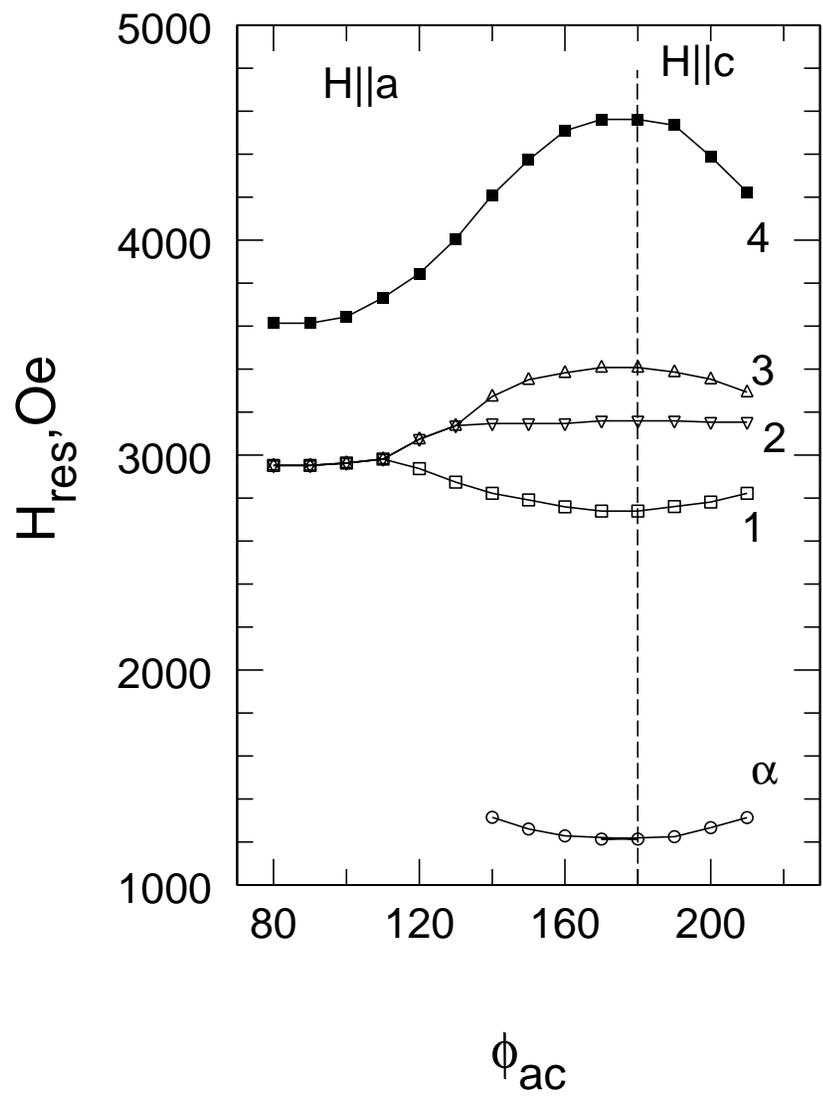
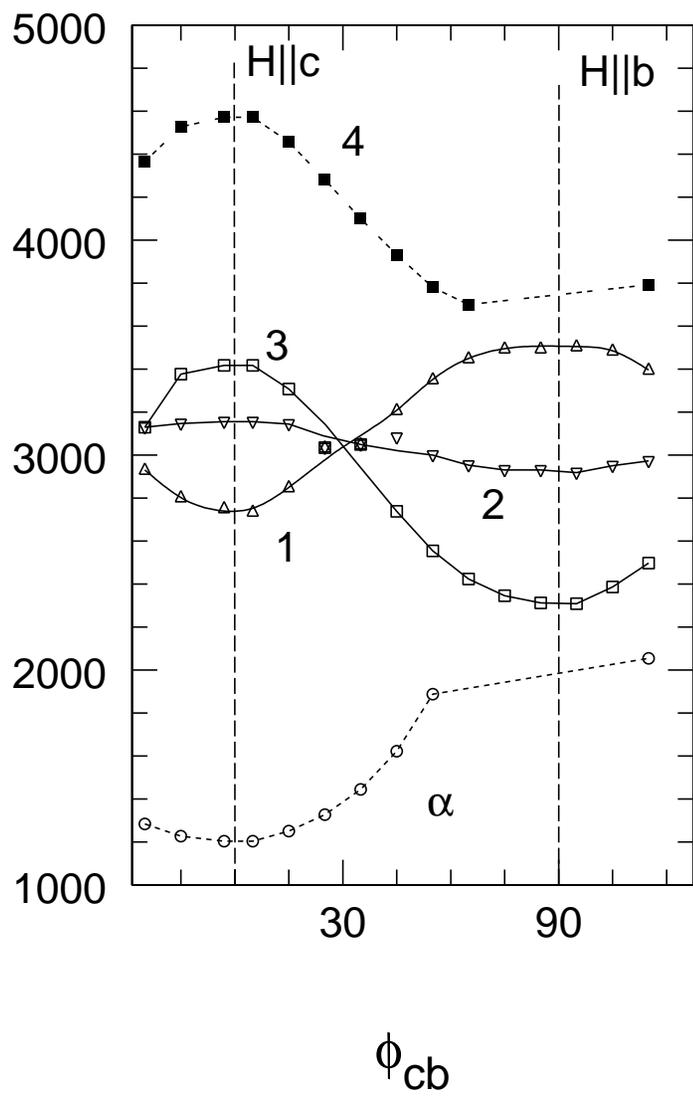
Fig.4. Angular dependencies of the ESR field at the frequency 9.4 GHz,
$T$=1.5~K for the rotation of the field in the planes $ac$ and $bc$.

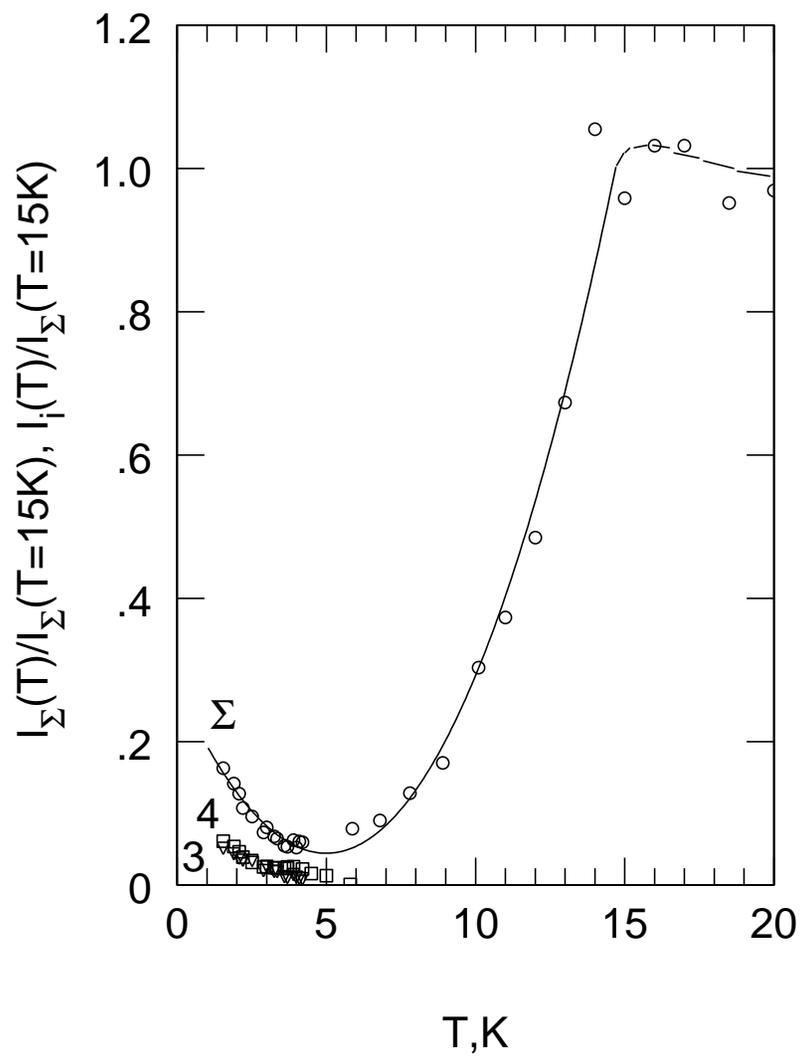
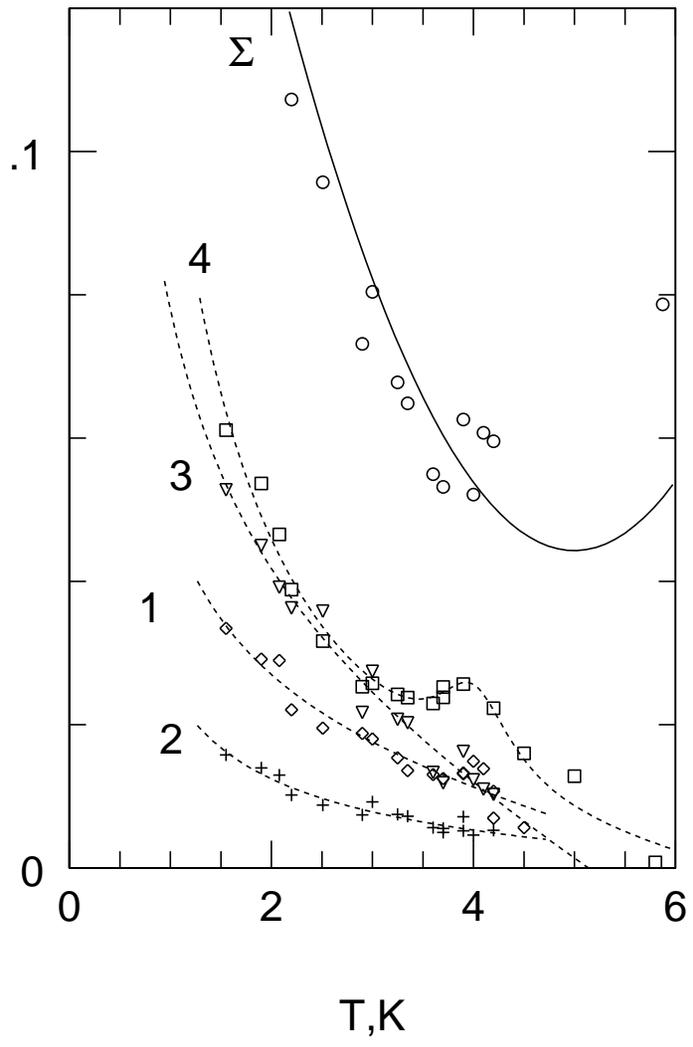
Fig.5. The dependence of the integral ESR intensity and of the
intensity of the lines 2,3,4 on the temperature  for the frequency 9.4
GHz. The numbers indicate the ESR lines according to
Fig.2., the integral intensity is marked by the sign $\sum$

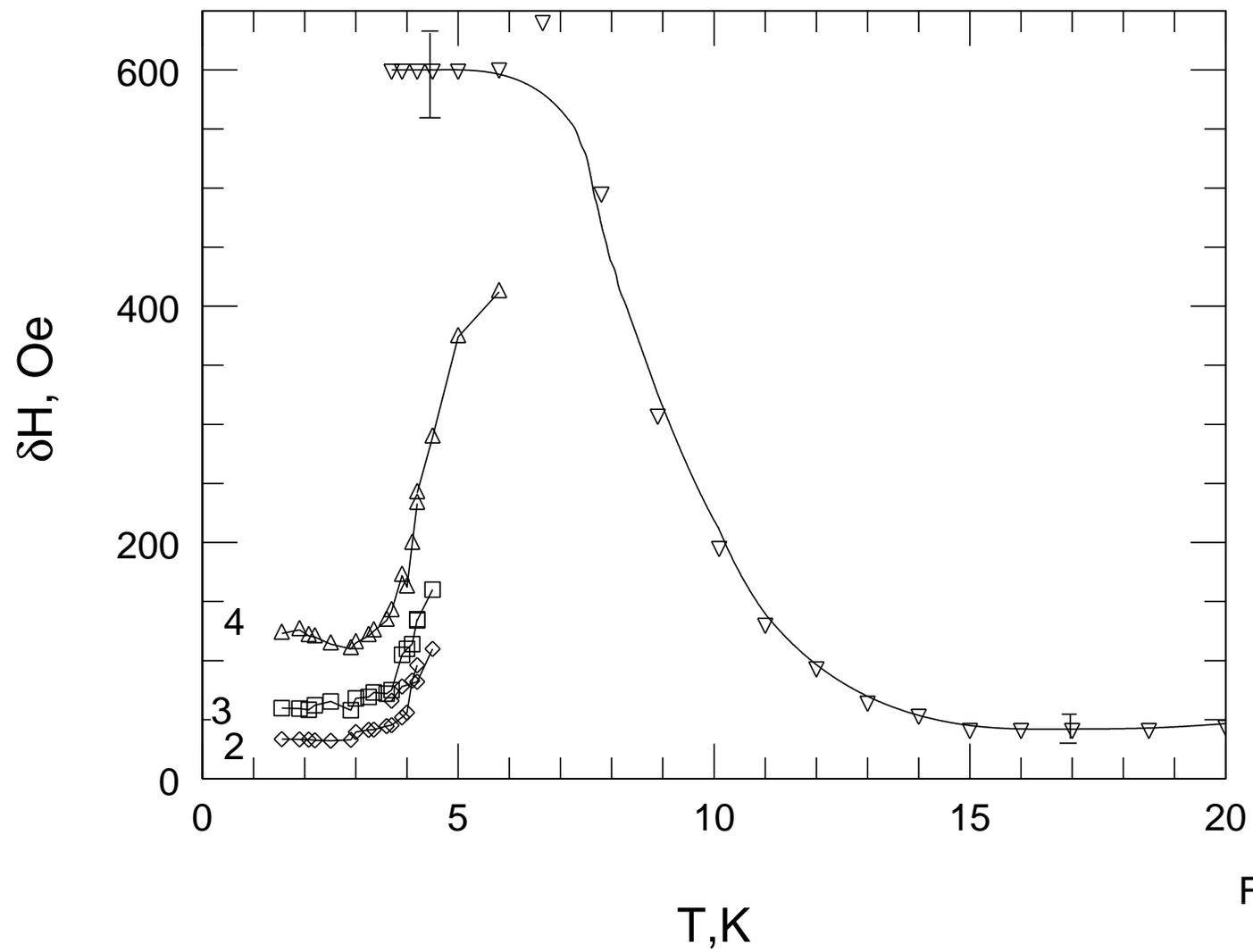
Fig.6. The dependence of the ESR linewidth on the temperature for the
frequency 9.4~GHz. The numbers are marking the resonant lines according
to Fig.2.

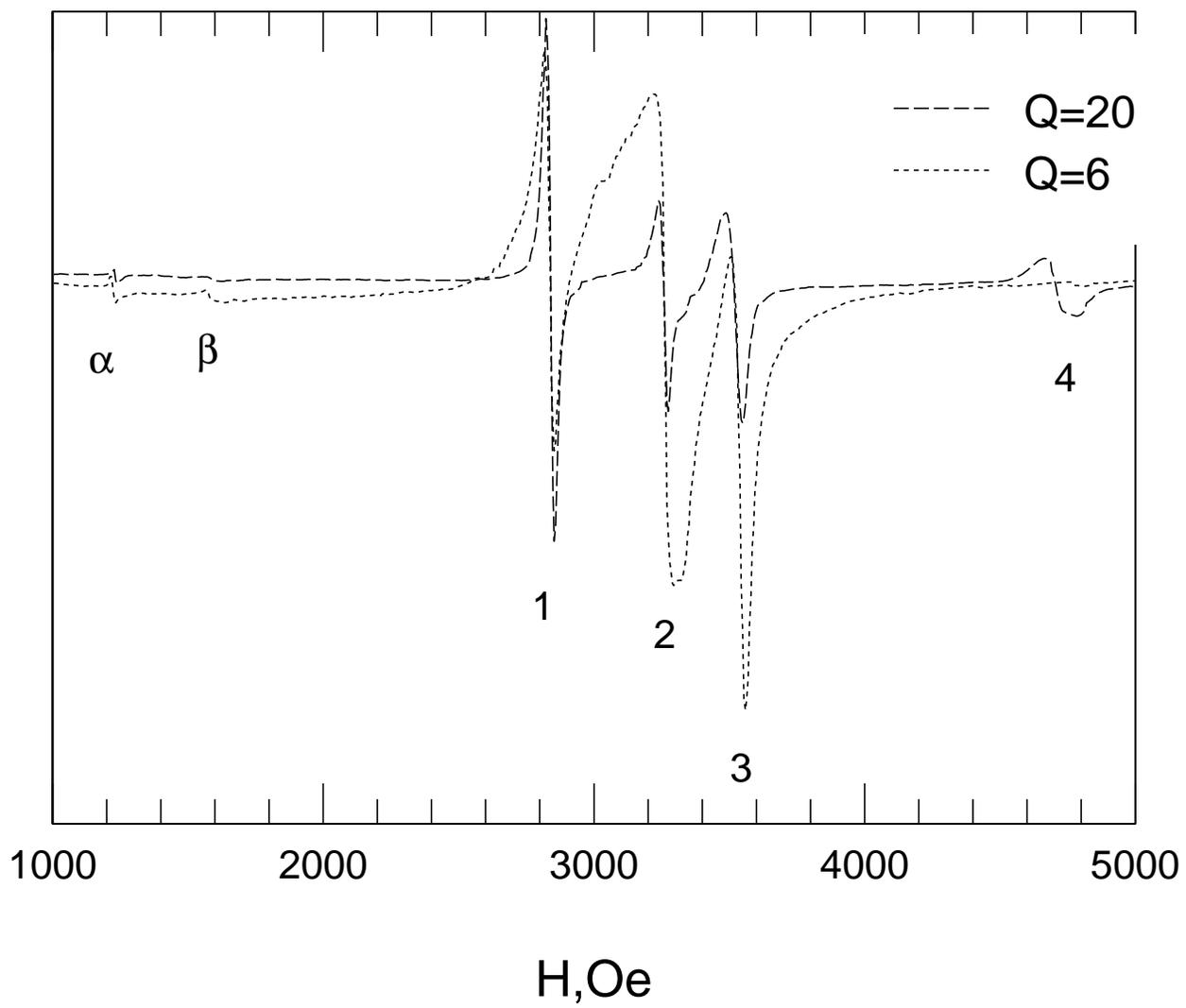
Fig.7. Comparison of the  absorption line derivatives of two samples
with $Q=$20 and $Q=$6 at $T=$1.5~K, on the frequency 9.1~GHz.
 The samples differ through the Fe-content and the growth
 methods. The amplitude of the signals is normalized to the equal
intensity at $T=$15K

Fig.8. The comparison of the ESR lines of the samples prepared from
the same components at the different crystallization rate. The lines
are taken at $T=1.8$~K, the  amplitude is normalized to equal
intensities at 15~K.

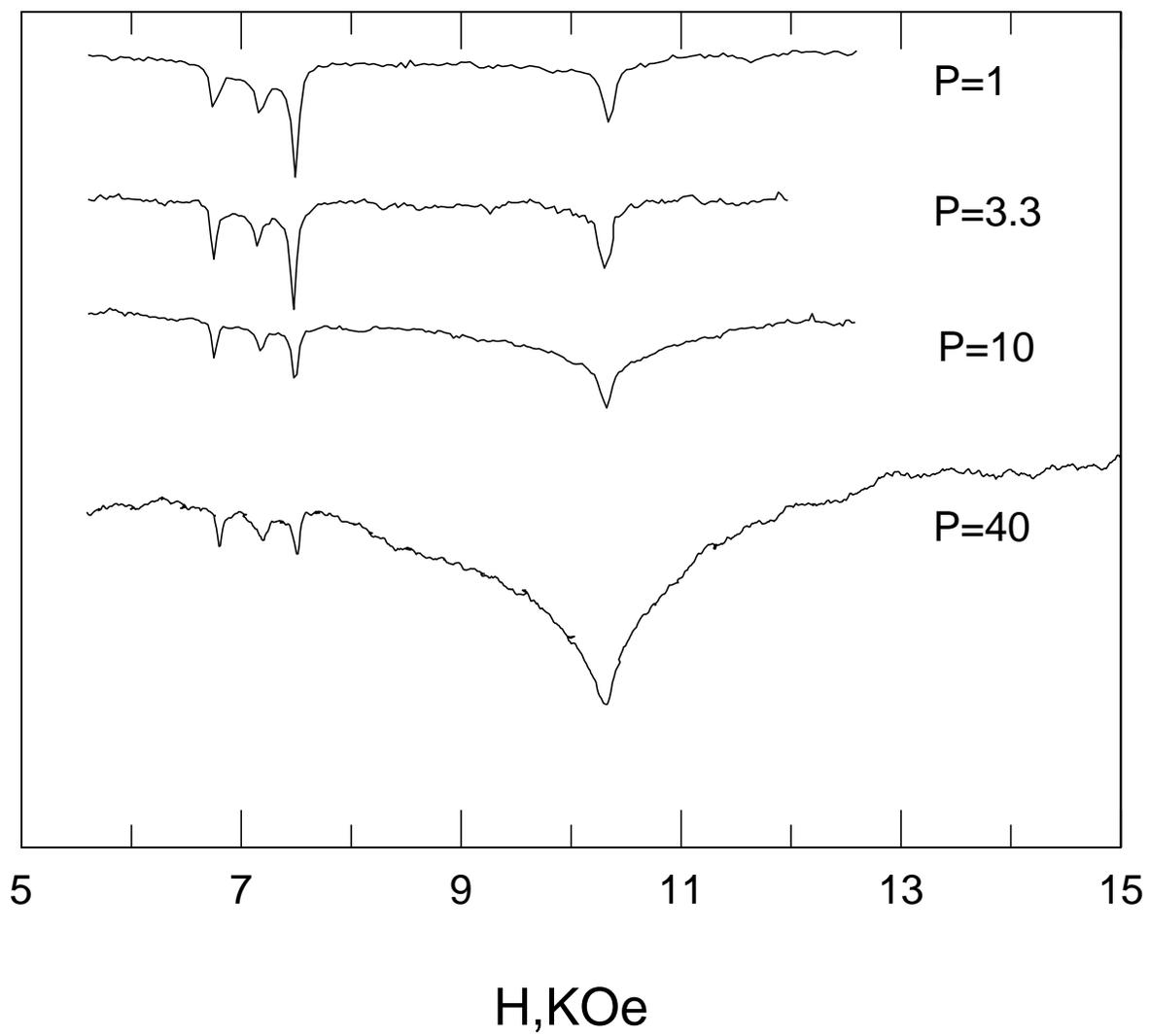
Fig.9. The field dependence of the power transmitted through the
resonator with the sample at the different microwave incident power
values, at the temperature 1.2-1.6~K.
${\bf H}
\parallel c$, $f=$20.2~GHz. The power values are given in arbitrary
units. The change of the temperature of the resonator is due to the
heating by microwave power.

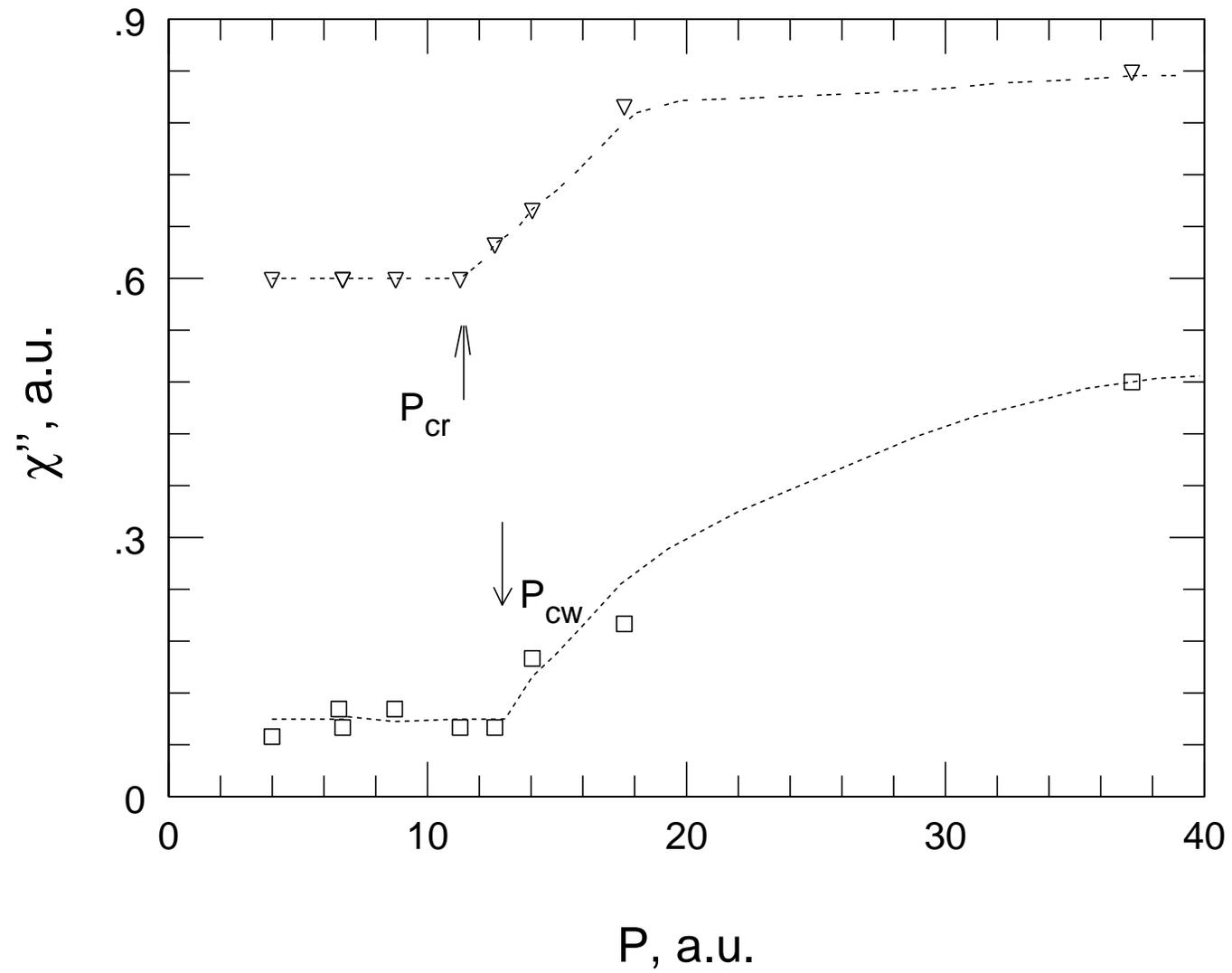
Fig.10. The dependence of the imaginary part of the magnetic
susceptibility on the microwave power for the ESR line 4 at the
resonant magnetic field (triangles) and at the left wing of the
resonant line (squares).
$P_{cr}$ and $P_{cw}$ are the threshold values of the power at the
resonance and on the wing correspondingly.
$T=$1.5~K.  ${\bf H}
\parallel c$, $f=$20.2~GHz

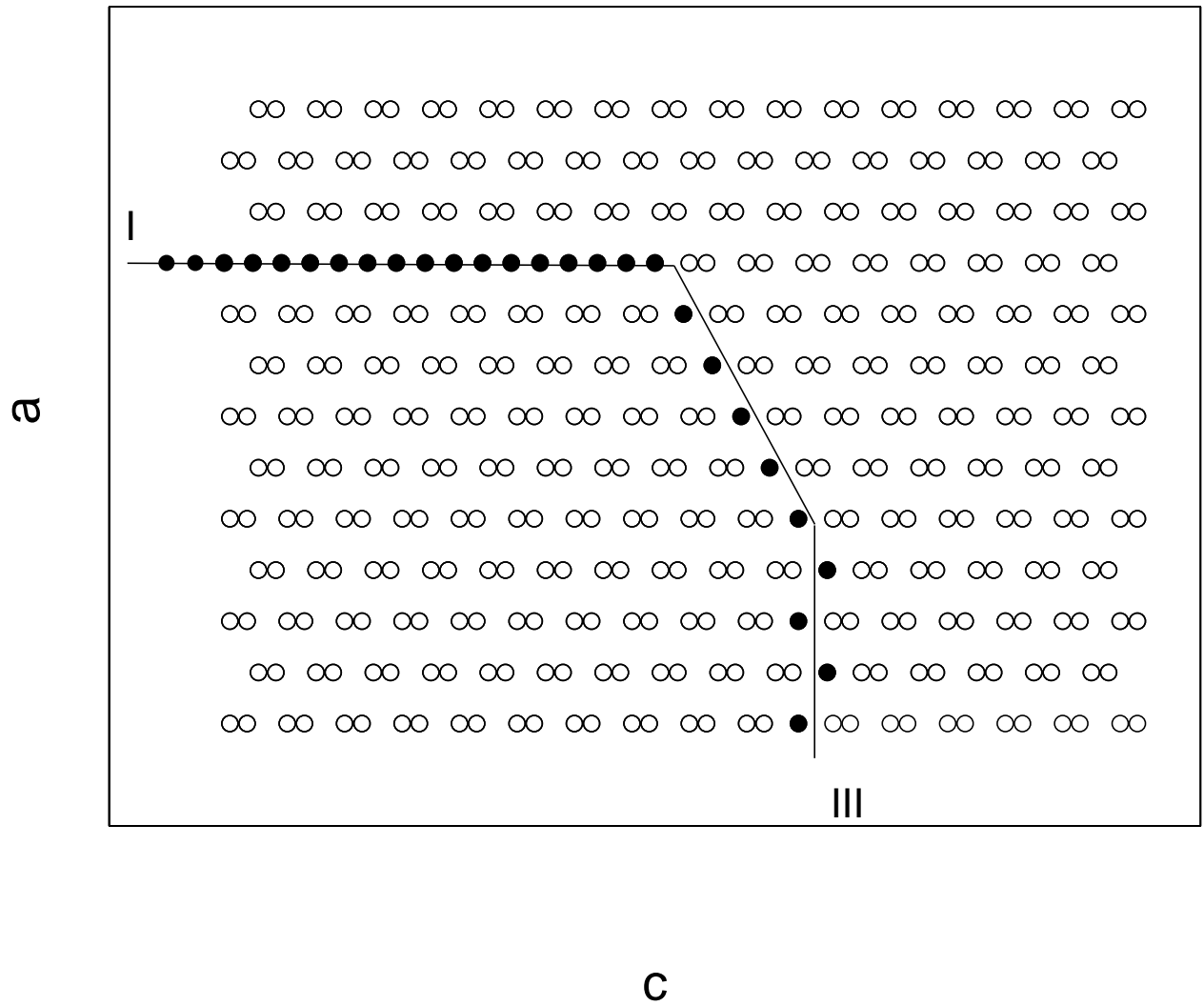
Fig.11. Scheme of the domains of the dimerization and of the domain
boundaries of types I and III. The filled circles are undimerized
Cu-ions.

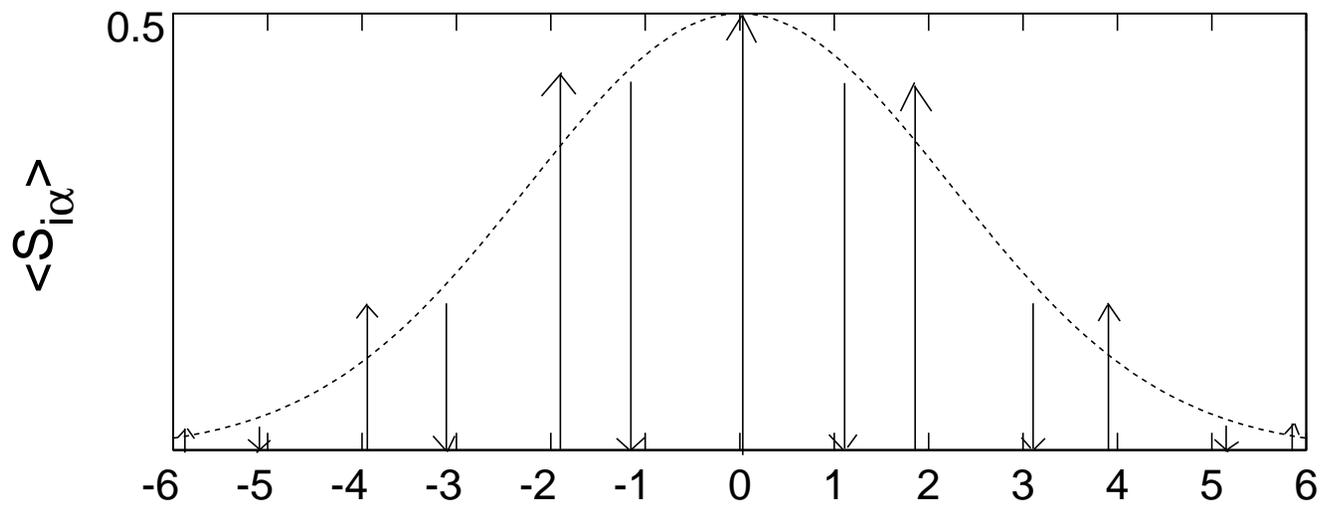
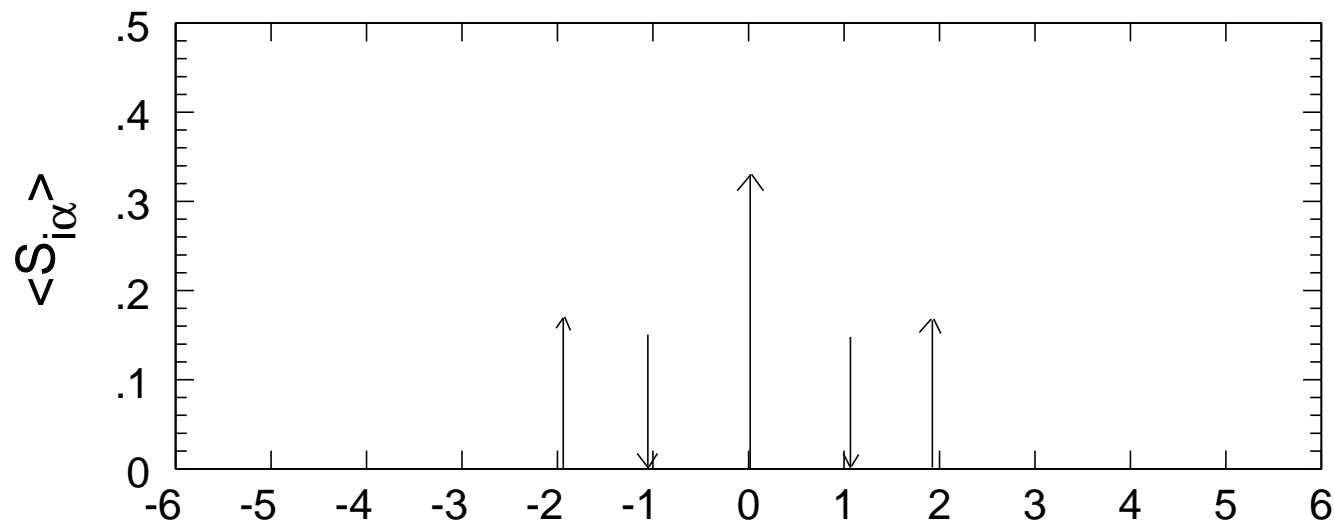
Fig.12. Proposed average spin projections in the vicinity of the
dimerization defect (upper part). Lower  part - calculated
average values of the spin projections for the five-spin-cluster
with the symmetric and antisymmetric exchange.


\begin{thebibliography}{90}

\bibitem{Hase}   M.Hase, I.Terasaki, K.Uchinokura,
Phys.Rev.Lett. {\bf 70} 3651 (1993)

\bibitem{Hirota} K.Hirota, D.E.Cox, J.E.Lorenzo, G.Shirane,
J.M.Tranquada, M.Hase, K.Uchinokura,
 H.Kojima, Y.Shibuya, I.Tanaka,
 Phys.Rev.Lett. {\bf 73} 736 (1994)

\bibitem{Regnault} L.-P.Regnault, M.A\"{i}n, B.Hennion, G.Dhalenne,
A.Revcholevschi, Phys.Rev.B {\bf 53} 5579 (1996)

\bibitem{Bray} J.W. Bray, H.Hart, L.Interrante, I.Jacobs, J.Kasper,
G.Watkins, S.Wee, J.Bonner,  Phys.Rev.Lett.  {\bf 35}, 744, (1975).

\bibitem{Bulaevsky}
L.N. Bulaevskii, Sov.Phys. Solid State {\bf 11}, 921 (1969)

\bibitem{Bethe} H.Bethe, Z.Phys {\bf 71}, 205 (1931)

\bibitem{Khomsky} D.Khomskii, W.Geertsma and M.Mostovoy,
Proceedings of the 21-st International
Conference on Low Temperature Physics,
Prague 1996. Chech. J. Phys.
{\bf 46} Suppl.S6 3239 (1996)

\bibitem{Coad} J.G.Lussier, S.M.Coad, D.F.McMorrow, D.Paul,
J.Phys.Condens.Matter {\bf 7} L325 (1995)

\bibitem{Regnault2}  L.-P.Regnault, J.-P.Renard, G.Dhallenne,
A.Revcholevschi,\\ Europhys.Lett. {\bf 32} 579 (1995)

\bibitem{Fukuyama} H.Fukuyama, T.Tanimoto, M.Saito,
J.Phys.Soc.Japan {\bf 65} 1182 (1996)

\bibitem{Honda} M.Honda, T.Shibata, K.Kindo, Sh. Sugai, T.Takeuchi,
H.Hori,  J.Phys.Soc.Jpn. {\bf 65} 691 (1996)

\bibitem{Smirnov1} A.I.Smirnov, V.N.Glazkov, A.N.Vasil'ev,
S.M.Coad, D.McK.Paul, G.Dhalenne, A.Revcholevschi, JETP
Lett.
{\bf 64} 305 (1996)

\bibitem{Smirnov2} A.I.Smirnov,  A.N.Vasil'ev, L.I.Leonyuk,
JETP Lett.
{\bf 64} 695 (1996)

\bibitem{Altshuler} S.A.Altshuler, B.M.Kosyrev, Electron
Paramagnetic Resonance,  Moscow,  "Nauka" 1972 (in russian)
 page 429

\bibitem{Chesnut1} M.T.Jones, D.B.Chesnut,  J.Chem.Phys. {\bf 38} 1311
(1963)

\bibitem{Chesnut2}
D.B.Chesnut, W.D.Philips, J.Chem.Phys. {\bf 35} 1002 (1961)

\bibitem{McConnell}  H.M.McConnell, H.O.Griffith, D.Pooley,
J.Chem.Phys {\bf 36} 2518 (1962)

\bibitem{Weiden} M.Weiden, W.Richter, C.Geibel, F.Steglich,
P.Lemmens, B.Eisener, M.Brinkman, G.Guntherodt, Physica B, {\bf 225}
177 (1996)

\bibitem{Fronzes} P.Fronzes, M.Poirier, A.Revcolevschi, G.Dhalenne
Phys.Rev.B {\bf 56}  7827 (1997)

\bibitem{Abragam1} A.Abragam, B.Bleaney, Electron Paramagnetic
Resonance of Transition Ions, v.1, "Clarendon Press", Oxford 1970,
		  chapter 1, section 5.

\bibitem{Abragam2} {\it ibid}, chapter 9 section 5.

\bibitem{Pisarev} D.N.Astrov, Sov.Phys.JETP {\bf 11} 708
1960

\bibitem{Jacobs}   I.S.Jacobs, J.W.Bray, H.R.Hart, Jr., L.V.Interrante,
J.S.Kasper, G.D.Watkins, D.E.Prober, J.C.Bonner Phys. Rev. B {\bf 14}
3036 (1976)

\bibitem{Ueda}   A.N.Vasil'ev, A.I.Smirnov, M.Isobe, Y.Ueda Ph.Rev.B
 {\bf 56} 5065 (1997)

\bibitem{Tsukerblat}  M.I.Belinskii, B.S.Tsukerblat,
A.V.Ablov Fiz.Tverd.Tela
{\bf 16} 989 (1974)   (Sov.Phys. Solid State)

\bibitem{Yablokov} Yu.V.Yablokov, V.K.Voronkova, L.V.Mosina, Electron
paramagnetic Resonance of the Exchange Clusters, Moscow, "Nauka",
1988\\
 Yu.V.Yablokov, B.Ya.Kuyavskaya, A.V.Ablov,
 L.V.Mosina, M.D.Mazus, Dokl.Akad.Nauk  {\bf 256} 1182 (1981)
 (Sov.Phys. Doklady)

\bibitem{Bazhan}
 B.S.Tsukerblat, V.M.Novotvorcev, B.Ya.Kuyavskaya, M.I.Belinskii
A.V.Ablov, A.N.Bazhan, V.T.Kalinnikov, Pisma Zh.Eksp.Teor.Fiz
{\bf 19} 525  (1974) (Sov.Phys. JETP Lett.)

\bibitem{Riera} J.Riera, A.Dobry, Ph.Rev.B {\bf 51}  16098 (1995)

\bibitem{Yamada} I.Yamada, M.Nishi, J.Akimitsu,  J.Phys.Cond Matt {\bf
8} 2625 (1996)

\bibitem{MuellerBonner}   G.Mueller, H.Thomas, H.Beck, J.C.Bonner,
		      Ph.Rev.B {\bf 24} 1429 (1981)

\bibitem{Ozhogin}  V.I.Ozhogin, Sov.Phys. JETP  {\bf 31} 1121 (1970)

\bibitem{Prozorova} V.V.Kveder, B.Ya.Kotyuzhanskii,
L.A.Prozorova, Sov.Phys.
JETP {\bf 47} 1165 (1973)

 \end{thebibliography}
\end{document}